\documentstyle[12pt,fleqn]{article}

\newlength{\dinwidth}
\newlength{\dinmargin}
\def\lapproxeq{\lower .7ex\hbox{$\;\stackrel{\textstyle<}{\sim}\;$}}
\def\gapproxeq{\lower .7ex\hbox{$\;\stackrel{\textstyle>}{\sim}\;$}}

\begin{document}
\begin{flushright}
MC-TH-98-10\\
June 1998\\
Revised: August 1998\\
\end{flushright}
\begin{center}
\vspace*{2cm}

{\Large \bf QCD coherence and jet rates in small $x$ deep inelastic scattering} \\

\vspace*{1cm}

J.R.~Forshaw and A.~Sabio Vera

\vspace*{0.5cm}
Department of Physics and Astronomy,\\
University of Manchester,\\
Manchester, M13 9PL, England.
\end{center}
\vspace*{5cm}
\begin{abstract}
   The contributions to the deep inelastic scattering structure function
which arise from emission of zero, one, two or three resolvable gluons and 
any number of unresolvable ones are computed to order 
${\bar \alpha}_{S}^{3}$. Coherence effects are taken into account 
via angular ordering and are demonstrated to yield (at the leading 
logarithm level) the identical results to those obtained assuming the 
multi-Regge kinematics of BFKL.
\end{abstract}
\newpage

\section{Introduction}
It is well known that the emission of soft gluons in perturbative QCD takes 
place into angular ordered regions $~\cite{C,DKMT,ESW,FR}$. This is called 
coherent emission. An important case in which soft gluons are involved is 
deep inelastic scattering (DIS) at small $x$. 

   For small enough values of Bjorken $x$ logarithms in $1/x$ need to be 
summed. This logarithmic summation is performed by the 
Balitsky-Fadin-Kuraev-Lipatov (BFKL) equation which at leading order sums 
terms $\sim [\alpha_{S}\ln (1/x)]^{n}$. Detailed discussions of the origin 
and derivation of the leading order BFKL equation can be found in~\cite{FR,
BFKL} and the next-to-leading order corrections can be found in $~\cite{NLOA,
NLOB}$. 

   The derivation of the BFKL equation relies upon the validity of the 
multi-Regge kinematics (i.e. strong ordering in the Sudakov variables). It 
turns out that this kinematic regime is generally only applicable for the 
calculation of elastic scattering and total cross-sections.
 
   For the calculation of more exclusive quantities, e.g. the number of 
gluons emitted in  deep inelastic scattering, we may well need to take into 
account QCD coherence effects, i.e. the use of the multi-Regge kinematics is 
no longer justified.

   In deep inelastic scattering, suppose the $(i-1){\rm th}$ emitted gluon 
(from the proton) has energy $E_{i-1}$ and that it emits a gluon with a 
fraction $(1-z_{i})$ of this energy and a transverse momentum of magnitude 
$q_{i}$. The (small) opening angle $\theta_{i}$ of this emitted gluon is 
given by
$$\theta_{i}\approx \frac{q_{i}}{(1-z_{i})E_{i-1}} ,$$
and $z_{i}$ is the fraction of the energy of the $(i-1){\rm th}$ gluon 
carried off by the $i{\rm th}$ gluon, i.e.
$$z_{i}=\frac{E_{i}}{E_{i-1}}.$$

   Colour coherence leads to angular ordering with increasing opening angles 
towards the hard scale (the photon) so in this case we have \(\theta_{i+1} 
> \theta_{i}\), which may be expressed as
$$\frac{q_{i+1}}{(1-z_{i+1})}>\frac{z_{i}q_{i}}{(1-z_{i})}.$$

   In the limit $z_{i},z_{i+1}\ll1$ this reduces to
$$q_{i+1}>z_{i}q_{i}.$$

   The kinematics of the virtual graphs (which reggeize the $t$-channel 
gluons) are similarly modified and ensure the cancellation of the collinear 
singularities in inclusive quantities.

   Before imposing the constraint of angular ordering, we first re-write the 
($t$=0) BFKL equation for $f_{\omega}(\mbox{\boldmath $k$})$, the 
unintegrated structure function in $\omega$-space ($\omega$ is the variable 
conjugate to $x$), in a form which will be suitable for the study of more 
exclusive quantities~\cite{C,M}:

$$f_{\omega}(\mbox{\boldmath $k$})=f_{\omega}^{0}(\mbox{\boldmath $k$})
+\bar\alpha_{S}\int\frac{d^{2}\mbox{\boldmath $q$}}{\pi q^{2}}\int_{0}^{1}
\frac{dz}{z}z^{\omega}\Delta_{R}(z,k)\Theta(q-\mu)f_{\omega}(\mbox{\boldmath 
$q$}+\mbox{\boldmath $k$}),$$
where $\mu$ is a collinear cutoff, \mbox{\boldmath $q$} is the transverse 
momentum of the emitted gluon, and the gluon Regge factor which sums all the 
virtual contributions is
$$\Delta_{R}(z_{i},k_{i})=\exp\left[-\bar\alpha_{S}\ln\frac{1}{z_{i}}\ln
\frac{k_{i}^{2}}{\mu^2}\right],$$
with \(k_i\equiv|\mbox{\boldmath $k$}_{i}|\), and $\bar\alpha_{S}\equiv 
C_{A}\alpha_{S}/\pi$, ($C_{A}=3$).

   The driving term, $f_{\omega}^{0}(\mbox{\boldmath $k$})$, includes the 
virtual corrections which reggeize the bare gluon. This form of the BFKL 
equation has a kernel which, under iteration, generates real gluon emissions 
with all the virtual corrections summed to all orders. As such, it is 
suitable for the study of the final state. Since $f_{\omega}$ is an inclusive 
structure function, it includes the sum over all final states and the 
$\mu$-dependence cancels between the real and virtual contributions.

   In this letter we wish to examine the individual contributions to the 
structure function of an on-shell gluon which come from the emission of $r$ 
gluons, each of which is constrained to have its transverse momentum less 
than $Q$ (where $\mu \ll Q$). By selecting an on-shell gluon as the target 
we can use the simple boundary condition

$$f_{\omega}^{0}(\mbox{\boldmath $k$})=\delta^{2}(\mbox{\boldmath $k$}).$$
Since the gluon is on shell it does not pick up any corrections due to 
reggeization. Note that our main conclusions do not depend upon the precise 
nature of the target particle.
  
   We define the structure function, $F_{0\omega}(Q,\mu)$, by integrating 
over all $\mu^{2} \leq q_{i}^{2}\leq Q^{2}$, i.e.
$$F_{0\omega}(Q,\mu) \equiv \Theta(Q-\mu) + \sum_{r=1}^{\infty}\int_{\mu^{2}}
^{Q^{2}}\prod_{i=1}^{r}\frac{d^{2}\mbox{\boldmath $q$}_{i}}{\pi q_{i}^{2}}
dz_{i}\frac{\bar{\alpha}_{S}}{z_{i}}z_{i}^{\omega}\Delta_{R}(z_{i},k_{i}), $$
and we have isolated the contributions from $i$ real gluon emissions by 
iterating the kernel explicitly.

   Consider the contributions to the structure function from a fixed number 
$r$ of emitted initial state gluons, $F_{0\omega}^{(r)}(Q)$, i.e.
$$F_{0\omega}(Q) = \int_{0}^{1} dx ~x^{\omega} F_{0}(x,Q) = 1 + 
\sum_{r=1}^{\infty} F_{0\omega}^{(r)}(Q).$$

   In this formulation (which does not include coherence) Marchesini~\cite{M} 
obtained the perturbative expansion for the $F_{0\omega}^{(r)}(Q,\mu)$. This 
is of the form
$$F^{(r)}_{0\omega}(Q,\mu)=\sum_{n=r}^{\infty}C^{(r)}_{0}(n;T)
\frac{\bar\alpha_{S}^{n}}{\omega^{n}},$$
with $T \equiv \ln({Q/\mu})$, and the inclusive structure function satisfies
$$F_{0\omega}(Q)\equiv\sum_{i=0}^{\infty}F_{0 \omega}^{(i)}(Q)=
\left(\frac{Q^2}{\mu^{2}}\right)^{\bar\gamma},$$
where $\bar\gamma$ is the BFKL anomalous dimension.

   Marchesini pointed out that coherence effects significantly modify the 
individual $F_{0\omega}^{(r)}(Q)$ whilst preserving the sum $F_{0\omega}(Q)$. 
He concluded that care must be taken to account properly for coherence in the 
calculation of associated distributions. 

   Modifying the BFKL formalism to account for coherence, 
$F_{0\omega}(Q,\mu)$ becomes
$$F_{\omega}(Q,\mu) = \Theta(Q-\mu) + \sum_{r=1}^{\infty}\int_{0}^{Q^{2}}
\prod_{i=1}^{r}\frac{d^{2}\mbox{\boldmath $q$}_{i}}{\pi q_{i}^{2}}dz_{i}
\frac{\bar{\alpha}_{S}}{z_{i}}z_{i}^{\omega}\Delta(z_{i},q_{i},k_{i})
\Theta(q_{i}-z_{i-1}q_{i-1}),$$
where $\Delta_{R}(z_{i},k_{i})$ is substituted by the coherence improved 
Regge factor 
$$\Delta(z_{i},q_{i},k_{i})=\exp\left[-\bar\alpha_{S}\ln\frac{1}{z_{i}}\ln
\frac{k_{i}^{2}}{z_{i}q_{i}^{2}}\right];~ ~ k_{i} > q_{i},$$
and for the first emission we take $q_{0}z_{0} = \mu$.

   The perturbative expansion of $F_{\omega}^{(r)}(Q)$ is now of the form
$$F_{\omega}^{(r)}(Q)=\sum_{n=r}^{\infty}\sum_{m=1}^{n}C^{(r)}(n,m;T)\frac{
\bar\alpha_{S}^{n}}{\omega^{2n-m}}.$$
   
   In the formalism with coherence no collinear cutoff is needed, except on 
the emission of the first gluon. This is because subsequent collinear 
emissions are regulated by the angular ordering constraint and it is those 
collinear emissions which induce the additional powers of $1/\omega$. 
Transforming to $x$-space it means that 

$$\frac{\bar \alpha_{S}^{n}}{\omega^{n+p}} \Longleftrightarrow \frac{\bar 
\alpha_{S}^{n}}{x} \left({\ln \frac{1}{x}}\right)^{n+p-1}, ~p<n,$$
i.e. coherence induces additional ${\ln (1/x)}$. In inclusive quantities the 
collinear singularities cancel. At a less inclusive level, such as for the 
associated distributions, the collinear singular terms need not cancel any 
more. 

\section{BFKL with a resolution scale}

   Although it is true that $~F^{(r)}_{0\omega}(Q) \neq F_{\omega}^{(r)}(Q)~$ 
we note that the $r$-gluon emission rate is not an observable quantity 
because in practise one can only detect emissions above some resolution 
scale, $\mu_{R}$. In this letter we intend to compute the $r$ resolved-gluon 
emission contributions to the structure function, i.e. we do not restrict 
the number of unresolved emissions which may occur.

   The experimental resolution scale $\mu_{R}$ is constrained by the 
collinear cutoff and the hard scale, $\mu \ll \mu_{R}\ll Q$. The 
implementation of a resolution scale in the BFKL equation has been studied 
by Lewis et al.~\cite{K}. In their work they derive a form of the BFKL 
equation which enables the structure of the gluon emissions to be studied 
in small $x$ deep inelastic scattering. The equation incorporates the 
summation of the virtual and unresolved real gluon emissions. They solve 
the equation to calculate the number of small $x$ deep inelastic events 
containing 0, 1, 2 ... resolved gluon jets.

We note that, within the leading log$(1/x)$ approximation,
 the resolved gluons can be 
identified as jets~\cite{K,MD} since any corrections arising from
additional radiation are suppressed by ${\cal O}(\alpha_s)$.
In this letter we are interested in the perturbative calculation,
to $\sim {\bar \alpha}_{S}^{3}$, of the $r$-jet cross-sections, 
where $r$ is the number of gluon emissions with transverse momentum bigger 
than $\mu_{R}$. 

   First we calculate the contribution from any number of emitted gluons with 
all of them unresolved. For the emission of a single unresolved gluon: 

\begin{eqnarray}
U&=& \int_{0}^{1}dz_{1}z_{1}^{\omega-1}\int_{\mu^{2}}^{\mu_{R}^{2}}
\frac{d^{2}\mbox{\boldmath $q$}_{1}}{\pi q_{1}^{2}}\left[\bar{\alpha}_{S} 
- \bar{\alpha}_{S}^{2}\ln\frac{1}{z_{1}}\ln\frac{q_{1}^{2}}{\mu^{2}} 
+ \frac{1}{2}\bar{\alpha}_{S}^{3}\ln^{2}\frac{1}{z_{1}}\ln^{2}
\frac{q_{1}^{2}}{\mu^{2}}\right] + ... \nonumber \\
&=&\frac{(2\bar{\alpha}_{S})}{\omega}S + \frac{(2\bar{\alpha}_{S})^{2}}
{\omega^{2}}\left[-\frac{1}{2}S^{2}\right] + \frac{(2\bar{\alpha}_{S})^{3}}
{\omega^{3}}\left[\frac{1}{3}S^{3}\right] + ...\end{eqnarray}

   For two unresolved emissions:

\begin{eqnarray}
UU &=& \int_{0}^{1}dz_{1}z_{1}^{\omega-1}\int_{0}^{1}dz_{2}z_{2}^{\omega-1}
\int_{\mu^{2}}^{\mu_{R}^{2}}\frac{d^{2}\mbox{\boldmath $q$}_{1}}
{\pi q_{1}^{2}}\int_{\mu^{2}}^{\mu_{R}^{2}}\frac{d^{2}
\mbox{\boldmath $q$}_{2}}{\pi q_{2}^{2}}\nonumber \\
& &\left[\bar{\alpha}_{S}^{2}-\bar{\alpha}_{S}^{3}\left(\ln\frac{1}{z_{1}}
\ln\frac{q_{1}^{2}}{\mu^2} + \ln\frac{1}{z_{2}}\ln\frac{k_{2}^{2}}{\mu^2}
\right)\right] + ... \nonumber \\
&=&\frac{(2\bar{\alpha}_{S})^{2}}{\omega^{2}}S^{2} + 
\frac{(2\bar{\alpha}_{S})^{3}}{\omega^{3}}\left[-\frac{7}{6}S^{3}\right] + ... 
\end{eqnarray}
where $\mbox{\boldmath $k$}_{i}=\mbox{\boldmath $k$}_{i-1}
-\mbox{\boldmath $q$}_{i}$, and we can write $\mbox{\boldmath $k$}^{2}_{i}
=[\sum_{n=1}^{i}\mbox{\boldmath $q$}_{n}]^{2}$. We have 
$\mbox{\boldmath $k$}_{0}=0$, and 
$$T \equiv \ln\frac{Q}{\mu_{R}},\;  S \equiv \ln\frac{\mu_{R}}{\mu}.$$

   The contribution from three unresolved emissions is

\begin{eqnarray}
UUU &=& \int_{0}^{1}dz_{1}z_{1}^{\omega-1}\int_{0}^{1}dz_{2}z_{2}^{\omega-1}
\int_{0}^{1}dz_{3}z_{3}^{\omega-1}\int_{\mu^{2}}^{\mu_{R}^{2}}\frac{d^{2}
\mbox{\boldmath $q$}_{1}}{\pi q_{1}^{2}}\int_{\mu^{2}}^{\mu_{R}^{2}}
\frac{d^{2}\mbox{\boldmath $q$}_{2}}{\pi q_{2}^{2}}
\int_{\mu^{2}}^{\mu_{R}^{2}}\frac{d^{2}\mbox{\boldmath $q$}_{3}}
{\pi q_{3}^{2}}\bar{\alpha}_{S}^{3} + ...  \nonumber \\
&=&\frac{(2\bar{\alpha}_{S})^{3}}{\omega^{3}}S^{3} + ... 
\end{eqnarray}

   Thus the 0-jet rate is

$$``{\rm 0-jet}"=U + UU + UUU + ... $$
\begin{eqnarray}
~~~~~~~~~~~~~~~~~~~~~~~~~=\frac{(2\bar{\alpha}_{S})}{\omega}S + 
\frac{(2\bar{\alpha}_{S})^{2}}{\omega^{2}}\left[\frac{S^{2}}{2}\right]
+\frac{(2\bar{\alpha}_{S})^{3}}{\omega^{3}}\left[\frac{S^{3}}{6}\right]+...
\end{eqnarray}

   Now we concentrate on calculating the 1-jet rate. For one resolved emission:
   
\begin{eqnarray}
R &=& \int_{0}^{1}dz_{1}z_{1}^{\omega-1}\int_{\mu_{R}^{2}}^{Q^{2}}
\frac{d^{2}\mbox{\boldmath $q$}_{1}}{\pi q_{1}^{2}}\left[\bar{\alpha}_{S} 
- \bar{\alpha}_{S}^{2}\ln\frac{1}{z_{1}}\ln\frac{q_{1}^{2}}{\mu^{2}} 
+ \frac{1}{2}\bar{\alpha}_{S}^{3}\ln^{2}\frac{1}{z_{1}}\ln^{2}
\frac{q_{1}^{2}}{\mu^{2}}\right] + ... \nonumber \\
&=& \frac{(2\bar{\alpha}_{S})}{\omega}T + \frac{(2\bar{\alpha}_{S})^{2}}
{\omega^{2}}\left[-\frac{1}{2}T^{2}-TS\right] + \frac{(2\bar{\alpha}_{S})
^{3}}{\omega^{3}}\left[\frac{1}{3}T^{3}+T^{2}S+TS^{2}\right] + ...  
\end{eqnarray}

   When the first emission is resolved and the second unresolved:
\begin{eqnarray}
RU &=& \int_{0}^{1}dz_{1}z_{1}^{\omega-1}\int_{0}^{1}dz_{2}z_{2}^{\omega-1}
\int_{\mu_{R}^{2}}^{Q^{2}}\frac{d^{2}\mbox{\boldmath $q$}_{1}}{\pi q_{1}^{2}}
\int_{\mu^{2}}^{\mu_{R}^{2}}\frac{d^{2}\mbox{\boldmath $q$}_{2}}
{\pi q_{2}^{2}}\nonumber \\
& &\left[\bar{\alpha}_{S}^{2}-\bar{\alpha}_{S}^{3}\left(\ln\frac{1}{z_{1}}
\ln\frac{q_{1}^{2}}{\mu^2} + \ln\frac{1}{z_{2}}\ln\frac{k_{2}^{2}}{\mu^2}
\right)\right] + ...  \nonumber \\
&=&  \frac{(2\bar{\alpha}_{S})^{2}}{\omega^{2}}TS + 
\frac{(2\bar{\alpha}_{S})^{3}}{\omega^{3}}\left[-T^{2}S-2TS^{2}\right] + ... 
\end{eqnarray}

   If the first emission is unresolved and the second resolved:

\begin{eqnarray}
UR&=& \int_{0}^{1}dz_{1}z_{1}^{\omega-1}\int_{0}^{1}dz_{2}z_{2}^{\omega-1}
\int_{\mu^{2}}^{\mu_{R}^{2}}\frac{d^{2}\mbox{\boldmath $q$}_{1}}
{\pi q_{1}^{2}}\int_{\mu_{R}^{2}}^{Q^{2}}\frac{d^{2}
\mbox{\boldmath $q$}_{2}}{\pi q_{2}^{2}}\nonumber \\
& &\left[\bar{\alpha}_{S}^{2}-\bar{\alpha}_{S}^{3}\left(\ln\frac{1}{z_{1}}
\ln\frac{q_{1}^{2}}{\mu^2} + \ln\frac{1}{z_{2}}\ln\frac{k_{2}^{2}}{\mu^2}
\right)\right] + ...  \nonumber \\
&=& \frac{(2\bar{\alpha}_{S})^{2}}{\omega^{2}}TS + 
\frac{(2\bar{\alpha}_{S})^{3}}{\omega^{3}}
\left[-\frac{3}{2}TS^{2}-\frac{1}{2}T^{2}S\right] + ... 
\end{eqnarray}
   
   Similarly for three emissions with two of them unresolved:
\begin{eqnarray} 
RUU &=& \int_{0}^{1}dz_{1}z_{1}^{\omega-1}\int_{0}^{1}dz_{2}z_{2}^{\omega-1}
\int_{0}^{1}dz_{3}z_{3}^{\omega-1}\int_{\mu_{R}^{2}}^{Q^{2}}\frac{d^{2}
\mbox{\boldmath $q$}_{1}}{\pi q_{1}^{2}}\int_{\mu^{2}}^{\mu_{R}^{2}}
\frac{d^{2}\mbox{\boldmath $q$}_{2}}{\pi q_{2}^{2}}\int_{\mu^{2}}
^{\mu_{R}^{2}}\frac{d^{2}\mbox{\boldmath $q$}_{3}}{\pi q_{3}^{2}}
\bar{\alpha}_{S}^{3} + ...  \nonumber \\
&=& \frac{(2\bar{\alpha}_{S})^{3}}{\omega^{3}}TS^{2} + ... 
\end{eqnarray}

\begin{eqnarray}
URU &=&\int_{0}^{1}dz_{1}z_{1}^{\omega-1}\int_{0}^{1}dz_{2}z_{2}
^{\omega-1}\int_{0}^{1}dz_{3}z_{3}^{\omega-1}\int_{\mu^{2}}^{\mu_{R}^{2}} 
\frac{d^{2}\mbox{\boldmath $q$}_{1}}{\pi q_{1}^{2}}\int_{\mu_{R}^{2}}^{Q^{2}}
\frac{d^{2}\mbox{\boldmath $q$}_{2}}{\pi q_{2}^{2}}\int_{\mu^{2}}
^{\mu_{R}^{2}}\frac{d^{2}\mbox{\boldmath $q$}_{3}}{\pi q_{3}^{2}}
\bar{\alpha}_{S}^{3} + ...  \nonumber \\
&=& \frac{(2\bar{\alpha}_{S})^{3}}{\omega^{3}}TS^{2} + ...  
\end{eqnarray}

\begin{eqnarray}
UUR &=& \int_{0}^{1}dz_{1}z_{1}^{\omega-1}\int_{0}^{1}dz_{2}z_{2}^{\omega-1}
\int_{0}^{1}dz_{3}z_{3}^{\omega-1}\int_{\mu^{2}}^{\mu_{R}^{2}} \frac{d^{2}
\mbox{\boldmath $q$}_{1}}{\pi q_{1}^{2}}\int_{\mu^{2}}^{\mu_{R}^{2}}
\frac{d^{2}\mbox{\boldmath $q$}_{2}}{\pi q_{2}^{2}}\int_{\mu_{R}^{2}}^{Q^{2}}
\frac{d^{2}\mbox{\boldmath $q$}_{3}}{\pi q_{3}^{2}}\bar{\alpha}_{S}^{3} 
+ ...  \nonumber \\
&=& \frac{(2\bar{\alpha}_{S})^{3}}{\omega^{3}}TS^{2} + ... 
\end{eqnarray}

The sum of these contributions is the 1-jet rate:

$$``{\rm 1-jet}" = R + RU + UR + RUU + URU + UUR + ... $$
\begin{eqnarray}
~~~~~=\frac{(2\bar{\alpha}_{S})}{\omega}T+\frac{(2\bar{\alpha}_{S})^{2}}
{\omega^{2}}\left[TS-\frac{1}{2}T^{2}\right]+\frac{(2\bar{\alpha}_{S})^{3}}
{\omega^{3}}\left[\frac{1}{3}T^{3}-\frac{1}{2}T^{2}S+\frac{1}{2}TS^{2}\right]
+...
\end{eqnarray}

   Let us now focus on the 2-jet rates, i.e. two of the emitted gluons have 
transverse momentum bigger than our resolution scale. There are several 
contributions, the first one comes from the case when only two gluons are 
emitted and both detected

\begin{eqnarray}
RR&=& \int_{0}^{1}dz_{1}z_{1}^{\omega-1}\int_{0}^{1}dz_{2}z_{2}^{\omega-1}
\int_{\mu_{R}^{2}}^{Q^{2}}\frac{d^{2}\mbox{\boldmath $q$}_{1}}{\pi q_{1}^{2}}
\int_{\mu_{R}^{2}}^{Q^{2}}\frac{d^{2}\mbox{\boldmath $q$}_{2}}{\pi q_{2}^{2}}
\nonumber \\
& &\left[\bar{\alpha}_{S}^{2}-\bar{\alpha}_{S}^{3}\left(\ln\frac{1}{z_{1}}
\ln\frac{q_{1}^{2}}{\mu^2} + \ln\frac{1}{z_{2}}\ln\frac{k_{2}^{2}}{\mu^2}
\right)\right] + ...  \nonumber \\
&=&\frac{(2\bar{\alpha}_{S})^{2}}{\omega^{2}}T^{2} + 
\frac{(2\bar{\alpha}_{S})^{3}}{\omega^{3}}\left[-\frac{7}{6}T^{3}
- 2 T^{2}S\right] + ... 
\end{eqnarray}
   
   If there is an additional undetected emission we must account for three more terms:

\begin{eqnarray}
RRU &=& \int_{0}^{1}dz_{1}z_{1}^{\omega-1}\int_{0}^{1}dz_{2}z_{2}^{\omega-1}
\int_{0}^{1}dz_{3}z_{3}^{\omega-1}\int_{\mu_{R}^{2}}^{Q^{2}}\frac{d^{2}
\mbox{\boldmath $q$}_{1}}{\pi q_{1}^{2}}\int_{\mu_{R}^{2}}^{Q^{2}}
\frac{d^{2}\mbox{\boldmath $q$}_{2}}{\pi q_{2}^{2}}\int_{\mu^{2}}
^{\mu_{R}^{2}}\frac{d^{2}\mbox{\boldmath $q$}_{3}}{\pi q_{3}^{2}}
\bar{\alpha}_{S}^{3} + ...  \nonumber \\
&=& \frac{(2\bar{\alpha}_{S})^{3}}{\omega^{3}}T^{2}S + ... 
\end{eqnarray}

\begin{eqnarray}
RUR &=& \int_{0}^{1}dz_{1}z_{1}^{\omega-1}\int_{0}^{1}dz_{2}z_{2}^{\omega-1}
\int_{0}^{1}dz_{3}z_{3}^{\omega-1}\int_{\mu_{R}^{2}}^{Q^{2}}\frac{d^{2}
\mbox{\boldmath $q$}_{1}}{\pi q_{1}^{2}}\int_{\mu^{2}}^{\mu_{R}^{2}}
\frac{d^{2}\mbox{\boldmath $q$}_{2}}{\pi q_{2}^{2}}\int_{\mu_{R}^{2}}^{Q^{2}}
\frac{d^{2}\mbox{\boldmath $q$}_{3}}{\pi q_{3}^{2}}\bar{\alpha}_{S}^{3}+... 
\nonumber \\
&=&\frac{(2\bar{\alpha}_{S})^{3}}{\omega^{3}}T^{2}S + ... 
\end{eqnarray}

\begin{eqnarray}
URR &=& \int_{0}^{1}dz_{1}z_{1}^{\omega-1}\int_{0}^{1}dz_{2}z_{2}^{\omega-1}
\int_{0}^{1}dz_{3}z_{3}^{\omega-1}\int_{\mu^{2}}^{\mu_{R}^{2}} \frac{d^{2}
\mbox{\boldmath $q$}_{1}}{\pi q_{1}^{2}}\int_{\mu_{R}^{2}}^{Q^{2}}\frac{d^{2}
\mbox{\boldmath $q$}_{2}}{\pi q_{2}^{2}}\int_{\mu_{R}^{2}}^{Q^{2}}\frac{d^{2}
\mbox{\boldmath $q$}_{3}}{\pi q_{3}^{2}}\bar{\alpha}_{S}^{3} + ...  
\nonumber \\
&=&\frac{(2\bar{\alpha}_{S})^{3}}{\omega^{3}}T^{2}S + ... 
\end{eqnarray}

and so

$$``{\rm 2-jet}"=RR + RRU + RUR + URR + ... $$
\begin{eqnarray}
~~~~~~~~~~~~~~~~~~~~~~=\frac{(2\bar{\alpha}_{S})^{2}}{\omega^{2}}\left[T^{2}
\right]+\frac{(2\bar{\alpha}_{S})^{3}}{\omega^{3}}\left[T^{2}S - \frac{7}{6}
T^{3} \right]+...
\end{eqnarray}

We now consider the emission of three resolved gluons. There is only one term
to order $\bar{\alpha}_{S}^{3}$, i.e.

\begin{eqnarray}
RRR &=& \int_{0}^{1}dz_{1}z_{1}^{\omega-1}\int_{0}^{1}dz_{2}z_{2}^{\omega-1}
\int_{0}^{1}dz_{3}z_{3}^{\omega-1}\int_ {\mu_{R}^{2}}^{Q^{2}}\frac{d^{2}
\mbox{\boldmath $q$}_{1}}{\pi q_{1}^{2}}\int_{\mu_{R}^{2}}^{Q^{2}}\frac{d^{2}
\mbox{\boldmath $q$}_{2}}{\pi q_{2}^{2}}\int_{\mu_{R}^{2}}^{Q^{2}}\frac{d^{2}
\mbox{\boldmath $q$}_{3}}{\pi q_{3}^{2}}\bar{\alpha}_{S}^{3} + ...  
\nonumber \\
&=&\frac{(2\bar{\alpha}_{S})^{3}}{\omega^{3}}T^{3} + ... 
= ``{\rm 3-jet}"
\end{eqnarray}

\section{Coherence with a resolution scale}

   Our aim in this section is to compute the 0-, 1-, 2-, 3-jet rates 
accounting for coherence. To proceed we must introduce the coherence 
condition $\Theta(q_{i}-z_{i-1}q_{i-1})$ and the coherence improved Regge 
factor, $\Delta(z_{i},q_{i},k_{i})$. For unresolved emissions (with the 
subscript ``$c$'' indicating coherence) we have

\begin{eqnarray}
U_{c} &=& \int_{0}^{1}dz_{1}z_{1}^{\omega-1}\int_{\mu^{2}}^{\mu_{R}^{2}}
\frac{d^{2}\mbox{\boldmath $q$}_{1}}{\pi q_{1}^{2}}\left[\bar{\alpha}_{S} 
- \bar{\alpha}_{S}^{2}\ln^{2}\frac{1}{z_{1}} + \frac{1}{2}\bar{\alpha}_{S}
^{3}\ln^{4}\frac{1}{z_{1}}\right] + ... \nonumber\\
   &=&\frac{(2\bar{\alpha}_{S})}{\omega}S + \frac{(2\bar{\alpha}_{S})^{2}}
{\omega^{2}}\left[-\frac{S}{\omega}\right] + \frac{(2\bar{\alpha}_{S})^{3}}
{\omega^{3}}\left[3\frac{S}{\omega^{2}}\right] + ... 
\end{eqnarray} 

\begin{eqnarray}
U_{c}U_{c} &=&\int_{0}^{1}dz_{1}z_{1}^{\omega-1}\int_{0}^{1}dz_{2}z_{2}
^{\omega-1}\int_{\mu^{2}}^{\mu_{R}^{2}}\frac{d^{2}\mbox{\boldmath $q$}_{1}}
{\pi q_{1}^{2}}
\int_{z_{1}^{2}q_{1}^{2}}^{\mu_{R}^{2}}\frac{d^{2}\mbox{\boldmath $q$}_{2}}
{\pi q_{2}^{2}}\nonumber\\ 
& &\left[\bar{\alpha}_{S}^{2}-\bar{\alpha}_{S}^{3}\left(\ln^{2}
\frac{1}{z_{1}}+\ln^{2}\frac{1}{z_{2}}+\ln\frac{1}{z_{2}}\ln
\frac{k_{2}^{2}}{q_{2}^{2}}\right)\right] + ...  \nonumber\\ 
&=&\frac{(2\bar{\alpha}_{S})^{2}}{\omega^{2}}\left[\frac{S}{\omega}
+\frac{S^{2}}{2}\right] + \frac{(2\bar{\alpha}_{S})^{3}}{\omega^{3}}
\left[-5\frac{S}{\omega^{2}} - \frac{S^{2}}{\omega}\right] + ... 
\end{eqnarray}

\begin{eqnarray}
U_{c}U_{c}U_{c} &=&\int_{0}^{1}dz_{1}z_{1}^{\omega-1}\int_{0}^{1}dz_{2}
z_{2}^{\omega-1}\int_{0}^{1}dz_{3}z_{3}^{\omega-1}\int_{\mu^{2}}
^{\mu_{R}^{2}}\frac{d^{2}\mbox{\boldmath $q$}_{1}}{\pi q_{1}^{2}}\int_{z_{1}
^{2}q_{1}^{2}}^{\mu_{R}^{2}}\frac{d^{2}\mbox{\boldmath $q$}_{2}}
{\pi q_{2}^{2}}\int_{z_{2}^{2}q_{2}^{2}}^{\mu_{R}^{2}}\frac{d^{2}
\mbox{\boldmath $q$}_{3}}{\pi q_{3}^{2}}\bar{\alpha}_{S}^{3}+...\nonumber\\ 
&=&\frac{(2\bar{\alpha}_{S})^{3}}{\omega^{3}}\left[2\frac{S}{\omega^{2}}
+\frac{S^{2}}{\omega}+\frac{1}{6}S^{3}\right]+...
\end{eqnarray}

   In the case of one single resolved emission, we have to consider 
(to order $\bar{\alpha}_{S}^{3}$) six terms:

\begin{eqnarray}
R_{c} &=& \int_{0}^{1}dz_{1}z_{1}^{\omega-1}\int_{\mu_{R}^{2}}^{Q^{2}}
\frac{d^{2}\mbox{\boldmath $q$}_{1}}{\pi q_{1}^{2}}\left[\bar{\alpha}_{S} 
- \bar{\alpha}_{S}^{2}\ln^{2}\frac{1}{z_{1}} + \frac{1}{2}
\bar{\alpha}_{S}^{3}\ln^{4}\frac{1}{z_{1}}\right] + ... \nonumber\\
   &=& \frac{(2\bar{\alpha}_{S})}{\omega}T + \frac{(2\bar{\alpha}_{S})^{2}}
{\omega^{2}}\left[-\frac{T}{\omega}\right] + \frac{(2\bar{\alpha}_{S})^{3}}
{\omega^{3}}\left[3\frac{T}{\omega^{2}}\right] + ... 
\end{eqnarray}

\begin{eqnarray}
R_{c}U_{c}&=& \int_{0}^{1}dz_{1}z_{1}^{\omega-1}\int_{0}^{1}dz_{2}z_{2}
^{\omega-1}\int_{\mu_{R}^{2}}^{Q^{2}}\frac{d^{2}\mbox{\boldmath $q$}_{1}}
{\pi q_{1}^{2}}\int_{z_{1}^{2}q_{1}^{2}}^{\mu_{R}^{2}}\frac{d^{2}
\mbox{\boldmath $q$}_{2}}{\pi q_{2}^{2}}
\Theta(\mu_{R}-z_{1}q_{1})\nonumber\\ 
& &\left[\bar{\alpha}_{S}^{2}-\bar{\alpha}_{S}^{3}\left(\ln^{2}
\frac{1}{z_{1}}+\ln^{2}\frac{1}{z_{2}}+\ln\frac{1}{z_{2}}\ln
\frac{k_{2}^{2}}{q_{2}^{2}}\right)\right] + ...  \nonumber\\ 
&=& \frac{(2\bar{\alpha}_{S})^{2}}{\omega^{2}}\left[\frac{T}{\omega} 
-\frac{1}{2}T^{2}\right] + \frac{(2\bar{\alpha}_{S})^{3}}{\omega^{3}}
\left[-5\frac{T}{\omega^{2}}+\frac{T^{2}}{\omega}\right] + ... 
\end{eqnarray}

\begin{eqnarray}
U_{c}R_{c}&=&\int_{0}^{1}dz_{1}z_{1}^{\omega-1}\int_{0}^{1}dz_{2}z_{2}
^{\omega-1}\int_{\mu^{2}}^{\mu_{R}^{2}}\frac{d^{2}\mbox{\boldmath $q$}_{1}}
{\pi q_{1}^{2}}\int_{\mu_{R}^{2}}^{Q^{2}}\frac{d^{2}\mbox{\boldmath $q$}_{2}}
{\pi q_{2}^{2}}\nonumber\\ 
& &\left[\bar{\alpha}_{S}^{2}-\bar{\alpha}_{S}^{3}\left(\ln^{2}\frac{1}{z_{1}}
+\ln^{2}\frac{1}{z_{2}}+\ln\frac{1}{z_{2}}\ln\frac{k_{2}^{2}}{q_{2}^{2}}
\right)\right] + ... \nonumber\\ 
&=&\frac{(2\bar{\alpha}_{S})^{2}}{\omega^{2}}TS + \frac{(2\bar{\alpha}_{S})
^{3}}{\omega^{3}}\left[-2\frac{TS}{\omega}\right] + ... 
\end{eqnarray}

\begin{eqnarray}
R_{c}U_{c}U_{c}&=&\int_{0}^{1}dz_{1}z_{1}^{\omega-1}\int_{0}^{1}dz_{2}z_{2}
^{\omega-1}\int_{0}^{1}dz_{3}z_{3}^{\omega-1}\int_{\mu_{R}^{2}}^{Q^{2}}
\frac{d^{2}\mbox{\boldmath $q$}_{1}}{\pi q_{1}^{2}}\int_{z_{1}^{2}q_{1}^{2}}
^{\mu_{R}^{2}}\frac{d^{2}\mbox{\boldmath $q$}_{2}}{\pi q_{2}^{2}}\int_{z_{2}
^{2}q_{2}^{2}}^{\mu_{R}^{2}}\frac{d^{2}\mbox{\boldmath $q$}_{3}}
{\pi q_{3}^{2}}\nonumber\\
& &\Theta(\mu_{R}-z_{1}q_{1})\bar{\alpha}_{S}^{3}+...\nonumber\\ 
&=&\frac{(2\bar{\alpha}_{S})^{3}}{\omega^{3}}\left[2\frac{T}{\omega^{2}}
-\frac{T^{2}}{\omega}+\frac{1}{3}T^{3}\right]+...
\end{eqnarray}

\begin{eqnarray}
U_{c}R_{c}U_{c} &=& \int_{0}^{1}dz_{1}z_{1}^{\omega-1}\int_{0}^{1}dz_{2}z_{2}
^{\omega-1}\int_{0}^{1}dz_{3}z_{3}^{\omega-1}\int_{\mu^{2}}^{\mu_{R}^{2}} 
\frac{d^{2}\mbox{\boldmath $q$}_{1}}{\pi q_{1}^{2}}\int_{\mu_{R}^{2}}^{Q^{2}}
\frac{d^{2}\mbox{\boldmath $q$}_{2}}{\pi q_{2}^{2}}\int_{z_{2}^{2}q_{2}^{2}}
^{\mu_{R}^{2}}\frac{d^{2}\mbox{\boldmath $q$}_{3}}{\pi q_{3}^{2}}\nonumber\\
& &\Theta(\mu_{R}-z_{2}q_{2})\bar{\alpha}_{S}^{3}+ ...\nonumber\\
&=&\frac{(2\bar{\alpha}_{S})^{3}}{\omega^{3}}\left[\frac{TS}{\omega}-\frac{1}
{2}T^{2}S\right]+...
\end{eqnarray}

\begin{eqnarray}
U_{c}U_{c}R_{c} &=& \int_{0}^{1}dz_{1}z_{1}^{\omega-1}\int_{0}^{1}dz_{2}z_{2}
^{\omega-1}\int_{0}^{1}dz_{3}z_{3}^{\omega-1}\int_{\mu^{2}}^{\mu_{R}^{2}} 
\frac{d^{2}\mbox{\boldmath $q$}_{1}}{\pi q_{1}^{2}}\int_{z_{1}^{2}q_{1}^{2}}
^{\mu_{R}^{2}}\frac{d^{2}\mbox{\boldmath $q$}_{2}}{\pi q_{2}^{2}}\int_{\mu_{R}
^{2}}^{Q^{2}}\frac{d^{2}\mbox{\boldmath $q$}_{3}}{\pi q_{3}^{2}}
\bar{\alpha}_{S}^{3} + ...\nonumber\\
&=&\frac{(2\bar{\alpha}_{S})^{3}}{\omega^{3}}\left[\frac{TS}{\omega}
+\frac{1}{2}TS^{2}\right]+...
\end{eqnarray}

   In these calculations we neglect terms which are beyond leading 
logarithmic approximation, i.e. terms suppressed by $\sim \omega^{n}$, with 
$(n \geq 1)$. 

   Now we consider the case when we resolve two of the emissions:

\begin{eqnarray}
R_{c}R_{c}&=& \int_{0}^{1}dz_{1}z_{1}^{\omega-1}\int_{0}^{1}dz_{2}z_{2}
^{\omega-1}\int_{\mu_{R}^{2}}^{Q^{2}}\frac{d^{2}\mbox{\boldmath $q$}_{1}}
{\pi q_{1}^{2}}\int_{max(\mu_{R}^{2},z_{1}^{2}q_{1}^{2})}^{Q^{2}}\frac{d^{2}
\mbox{\boldmath $q$}_{2}}{\pi q_{2}^{2}}\nonumber\\ 
& &\left[\bar{\alpha}_{S}^{2}-\bar{\alpha}_{S}^{3}\left(\ln^{2}\frac{1}{z_{1}}
+\ln^{2}\frac{1}{z_{2}}+\ln\frac{1}{z_{2}}\ln\frac{k_{2}^{2}}{q_{2}^{2}}
\right)\right] + ... \nonumber\\ 
&=&\frac{(2\bar{\alpha}_{S})^{2}}{\omega^{2}}\left[T^{2}\right] 
+ \frac{(2\bar{\alpha}_{S})^{3}}{\omega^{3}}\left[-2\frac{T^{2}}{\omega}
\right] + ... 
\end{eqnarray}

\begin{eqnarray}
R_{c}R_{c}U_{c} &=&\int_{0}^{1}dz_{1}z_{1}^{\omega-1}\int_{0}^{1}dz_{2}z_{2}
^{\omega-1}\int_{0}^{1}dz_{3}z_{3}^{\omega-1}\int_{\mu_{R}^{2}}^{Q^{2}}
\frac{d^{2}\mbox{\boldmath $q$}_{1}}{\pi q_{1}^{2}}\int_{max(\mu_{R}^{2},
z_{1}^{2}q_{1}^{2})}^{Q^{2}}\frac{d^{2}\mbox{\boldmath $q$}_{2}}{\pi q_{2}
^{2}}\int_{z_{2}^{2}q_{2}^{2}}^{\mu_{R}^{2}}\frac{d^{2}
\mbox{\boldmath $q$}_{3}}{\pi q_{3}^{2}}\nonumber\\
& &\Theta(\mu_{R}-z_{2}q_{2})\bar{\alpha}_{S}^{3}+...\nonumber\\ 
&=&\frac{(2\bar{\alpha}_{S})^{3}}{\omega^{3}}\left[\frac{T^{2}}
{\omega}-\frac{2}{3}T^{3}\right]+...
\end{eqnarray}

\begin{eqnarray}
R_{c}U_{c}R_{c} &=&\int_{0}^{1}dz_{1}z_{1}^{\omega-1}\int_{0}^{1}dz_{2}z_{2}
^{\omega-1}\int_{0}^{1}dz_{3}z_{3}^{\omega-1}\int_{\mu_{R}^{2}}^{Q^{2}}
\frac{d^{2}\mbox{\boldmath $q$}_{1}}{\pi q_{1}^{2}}\int_{z_{1}^{2}q_{1}^{2}}
^{\mu_{R}^{2}}\frac{d^{2}\mbox{\boldmath $q$}_{2}}{\pi q_{2}^{2}}\int_{\mu_{R}
^{2}}^{Q^{2}}\frac{d^{2}\mbox{\boldmath $q$}_{3}}{\pi q_{3}^{2}}\nonumber\\
& &\Theta(\mu_{R}-z_{1}q_{1})\bar{\alpha}_{S}^{3}+...\nonumber\\ 
&=&\frac{(2\bar{\alpha}_{S})^{3}}{\omega^{3}}\left[\frac{T^{2}}{\omega}
-\frac{T^{3}}{2}\right]+...
\end{eqnarray}

\begin{eqnarray}
U_{c}R_{c}R_{c} &=&\int_{0}^{1}dz_{1}z_{1}^{\omega-1}\int_{0}^{1}dz_{2}z_{2}
^{\omega-1}\int_{0}^{1}dz_{3}z_{3}^{\omega-1}\int_{\mu^{2}}^{\mu_{R}^{2}}
\frac{d^{2}\mbox{\boldmath $q$}_{1}}{\pi q_{1}^{2}}\int_{\mu_{R}^{2}}^{Q^{2}}
\frac{d^{2}\mbox{\boldmath $q$}_{2}}{\pi q_{2}^{2}}\int_{max(\mu_{R}^{2},z_{2}
^{2}q_{2}^{2})}^{Q^{2}}\frac{d^{2}\mbox{\boldmath $q$}_{3}}{\pi q_{3}^{2}}
\nonumber\\
& &\bar{\alpha}_{S}^{3}+...\nonumber\\ 
&=&\frac{(2\bar{\alpha}_{S})^{3}}{\omega^{3}}\left[T^{2}S\right]+...
\end{eqnarray}

   Finally, if we have three resolved emissions then

\begin{eqnarray}
R_{c}R_{c}R_{c} &=&\int_{0}^{1}dz_{1}z_{1}^{\omega-1}\int_{0}^{1}dz_{2}z_{2}
^{\omega-1}\int_{0}^{1}dz_{3}z_{3}^{\omega-1}\int_{\mu_{R}^{2}}^{Q^{2}}
\frac{d^{2}\mbox{\boldmath $q$}_{1}}{\pi q_{1}^{2}}\int_{max(\mu_{R}^{2},z_{1}
^{2}q_{1}^{2})}^{Q^{2}}\frac{d^{2}\mbox{\boldmath $q$}_{2}}{\pi q_{2}^{2}}
\int_{max(\mu_{R}^{2},z_{2}^{2}q_{2}^{2})}^{Q^{2}}\frac{d^{2}\mbox{\boldmath 
$q$}_{3}}{\pi q_{3}^{2}}\nonumber\\
& &\bar{\alpha}_{S}^{3}+...\nonumber\\ 
&=&\frac{(2\bar{\alpha}_{S})^{3}}{\omega^{3}}\left[T^{3}\right]+...
\end{eqnarray}

   At first sight these results are completely different from those
computed without coherence (BFKL). It is noteworthy that there exist stronger 
singularities ($\omega \rightarrow 0$) than occur in the BFKL approach. 
The presence of these new singularities 
may lead one to think that the calculation of exclusive quantities with the 
BFKL equation is destined to give incorrect expressions, and that the correct 
solution to the problem is to introduce coherence. However, if we calculate
 the 0-, 1-, 2-, 3-jet production rates with coherence we obtain the following
expressions.

$$``{\rm 0-jet}" =U_{c}+U_{c}U_{c}+U_{c}U_{c}U_{c}+...$$
\begin{eqnarray}
~~~~~~~~~~~~~~~~~~~~~~~=\frac{(2\bar{\alpha}_{S})}{\omega}S
+\frac{(2\bar{\alpha}_{S})^{2}}{\omega^{2}}\left[\frac{S^{2}}{2}\right]
+\frac{(2\bar{\alpha}_{S})^{3}}{\omega^{3}}\left[\frac{S^{3}}{6}\right]
+...\end{eqnarray}   

$$``{\rm 1-jet}" = R_{c}+R_{c}U_{c}+U_{c}R_{c}+R_{c}U_{c}U_{c}+U_{c}R_{c}U_{c}
+U_{c}U_{c}R_{c}+...$$
\begin{eqnarray}
~~~~~~=\frac{(2\bar{\alpha}_{S})}{\omega}T+\frac{(2\bar{\alpha}_{S})^{2}}
{\omega^{2}}\left[TS-\frac{1}{2}T^{2}\right]+\frac{(2\bar{\alpha}_{S})^{3}}
{\omega^{3}}\left[\frac{1}{3}T^{3}-\frac{1}{2}T^{2}S+\frac{1}{2}TS^{2}\right]
+...\end{eqnarray}

$$``{\rm 2-jet}"= R_{c}R_{c}+R_{c}R_{c}U_{c}+R_{c}U_{c}R_{c}+U_{c}R_{c}R_{c}
+...$$
\begin{eqnarray}
~~~~~~~~~~~~~~~~~~~~~~~~~~=\frac{(2\bar{\alpha}_{S})^{2}}{\omega^{2}}
\left[T^{2}\right]+\frac{(2\bar{\alpha}_{S})^{3}}{\omega^{3}}
\left[T^{2}S-\frac{7}{6}T^{3}\right]+...\end{eqnarray}

$$``{\rm 3-jet}"= R_{c}R_{c}R_{c}+...$$
\begin{eqnarray}
~~~~~~~~~~~~~~~~~~~~~~~~~~~~~~~~~~~~~~~~~~~=\frac{(2\bar{\alpha}_{S})^{3}}
{\omega^{3}}\left[T^{3}\right]+...\end{eqnarray}

   Note that the additional ``coherence induced'' logarithms cancel and that 
these results are identical to those obtained without coherence, i.e. (4, 11, 
16, 17). Presumably this cancellation persists for $n$-jet rates to all 
orders in ${\bar \alpha}_{S}$.
   
\section{Conclusions}

\begin{table}[h] 
    \begin{tabular}{|c|l|c|l|c|l|c|l|}  \hline 
    \multicolumn{8}{|c|}{Table 1: BFKL}\\ \hline\hline 
    $F^{(1)}_{0\omega}$  =&$U$&+&$R$&&&&\\
    $F^{(2)}_{0\omega}$  =&$UU$&+&$RU+UR$&+&$RR$&& \\
    $F^{(3)}_{0\omega}$  =&$UUU$&+&$RUU+URU+UUR$&+&$RRU+RUR+URR$&+&$RRR$ \\ 
\hline
    &0-jet&&1-jet&&2-jet&&3-jet\\ \hline
    \end{tabular} 
\end{table}

On summing the $n{\rm th}$ row in Table 1 the dependence on the resolution 
scale disappears and leads to $F_{0 \omega}^{(n)}$ which is different from 
the $F_{\omega}^{(n)}$ computed by summing the corresponding row of Table 2. 
This is the result demonstrated in $~\cite{M}$, i.e. for BFKL

$$U + R = $$
\begin{eqnarray}
~~\frac{(2\bar{\alpha}_{S})}{\omega}\left[T+S\right]+\frac{(2\bar{\alpha}_{S})
^{2}}{\omega^{2}}\left[-\frac{1}{2}(T+S)^{2}\right]+\frac{(2\bar{\alpha}_{S})
^{3}}{\omega^{3}}\left[\frac{1}{3}(T+S)^{3}\right]+...= F_{0,\omega}^{(1)}(Q)
\end{eqnarray}

$$UU + RU + UR + RR = $$
\begin{eqnarray}
~~~~~~~~~~~~~~~~~~~~\frac{(2\bar{\alpha}_{S})^{2}}{\omega^{2}}\left[(T+S)^{2}
\right]+\frac{(2\bar{\alpha}_{S})^{3}}{\omega^{3}}\left[-\frac{7}{6}(T+S)
^{3} \right]+...= F_{0,\omega}^{(2)}(Q)
\end{eqnarray}

$$UUU + RUU + URU + UUR + RRU + RUR + URR + RRR = $$
\begin{eqnarray}
~~~~~~~~~~~~~~~~~~~~~~~~~~~~~~~~~~~\frac{(2\bar{\alpha}_{S})^{3}}
{\omega^{3}}\left[(T+S)^{3}\right]+...= F_{0,\omega}^{(3)}(Q)
\end{eqnarray}

\begin{table}[h] 
    \begin{tabular}{|c|l|c|l|c|l|c|l|}  \hline
    \multicolumn{8}{|c|}{Table 2: COHERENCE}\\ \hline\hline  
    $F^{(1)}_{\omega}$  =&$U_{c}$&+&$R_{c}$&&&&\\
    $F^{(2)}_{\omega}$  =&$U_{c}U_{c}$&+&$R_{c}U_{c}+U_{c}R_{c}$&+&$R_{c}
R_{c}$&& \\
    $F^{(3)}_{\omega}$  =&$U_{c}U_{c}U_{c}$&+&$R_{c}U_{c}U_{c}+U_{c}R_{c}
U_{c}+U_{c}U_{c}R_{c}$&+&$R_{c}R_{c}U_{c}+R_{c}U_{c}R_{c}+U_{c}R_{c}R_{c}$&
+&$R_{c}R_{c}R_{c}$ \\ \hline
    &0-jet&&1-jet&&2-jet&&3-jet\\ \hline
    \end{tabular}
\end{table}

Whilst for the terms with coherence one finds

$$U_{c} + R_{c}=$$
\begin{eqnarray}
~~\frac{(2\bar{\alpha}_{S})}{\omega}\left[T+S\right]+\frac{(2\bar{\alpha}_{S})
^{2}}{\omega^{2}}\left[-\frac{1}{\omega}(T+S)\right]+\frac{(2\bar{\alpha}_{S})
^{3}}{\omega^{3}}\left[\frac{3}{\omega^{2}}(T+S)\right]+...= 
F_{\omega}^{(1)}(Q)\end{eqnarray}

$$U_{c}U_{c} + R_{c}U_{c} + U_{c}R_{c} +R_{c}R_{c}=$$
\begin{eqnarray}
\frac{(2\bar{\alpha}_{S})^{2}}{\omega^{2}}\left[\frac{1}{2}(T+S)^{2}
+\frac{1}{\omega}(T+S)\right]+\frac{(2\bar{\alpha}_{S})^{3}}{\omega^{3}}
\left[-\frac{1}{\omega}(T+S)^{2}-\frac{5}{\omega^{2}}(T+S)\right]+...= 
F_{\omega}^{(2)}(Q)\end{eqnarray}

$$U_{c}U_{c}U_{c} + R_{c}U_{c}U_{c} + U_{c}R_{c}U_{c} + U_{c}U_{c}R_{c} + 
R_{c}R_{c}U_{c} + R_{c}U_{c}R_{c}+ U_{c}R_{c}R_{c} + R_{c}R_{c}R_{c}=$$
\begin{eqnarray}
~~~~~~~~\frac{(2\bar{\alpha}_{S})^{3}}{\omega^{3}}\left[\frac{1}{6}(T+S)^{3}
+\frac{1}{\omega}(T+S)^{2}+\frac{2}{\omega^{2}}(T+S)\right]+...= F_{\omega}
^{(3)}(Q)\end{eqnarray}

   However, summing the columns in each table, one obtains the more physical 
$n$-jet rates. In this case the BFKL and coherence results coincide. We note 
that this cancellation supports the work of $~\cite{K,SO}$.

   We have shown the explicit cancellation of coherence induced collinear 
singularities in $n$-jet rate calculations to order ${\bar \alpha}_{S}^{3}$ 
at the leading logarithm level. Nevertheless we wish to remark that this is 
not to say that coherence effects are always unimportant. In particular, we 
have neglected formally subleading terms in $F_{\omega}^{(r)}$ which are only 
suppressed by factors $\sim (\omega T)^{n}$ (with $n>0$). Those terms are 
relevant if we go beyond the leading ${\ln (1/x)}$ approximation.

\section*{Acknowledgements}

   We thank Gavin Salam and Graham Shaw for useful comments.

\end{document}